\documentclass[12pt,a4paper]{amsart}
\usepackage{amsfonts}
\usepackage{amsthm}
\usepackage{amsmath}
\usepackage{amssymb}
\usepackage{enumitem,kantlipsum}
\usepackage{graphicx}

\usepackage[colorlinks=true, allcolors=black]{hyperref}

\theoremstyle{definition}

\theoremstyle{remark}

\numberwithin{equation}{section}

\title[Marchuk's models of infection diseases: new developments]{Marchuk's models of infection diseases: new developments}
\author{Irina Volinsky, Alexander Domoshnitsky, Marina Bershadsky and Roman Shklyar}
\address{Ariel university, Ariel 40700, Israel\\ Department of Mathematics}
\email {irinav@ariel.ac.il}
\email {adom@ariel.ac.il}

\begin{document}

\begin{abstract}
We consider mathematical models of infection diseases built by G.I. Marchuk in his well known book on immunology. These models are in the form of systems of ordinary delay differential equations. We add a distributed control in one of the equations describing the dynamics of the antibody concentration rate. Distributed control looks here naturally since the change of this concentration rather depends on the corresponding average value of the difference of the current and normal antibody concentrations on the time interval than on their difference at the point $t$ only.  
Choosing this control in a corresponding form, we propose some ideas of the stabilization in the cases, where other methods do not work. The main idea is to reduce the stability analysis of a given integro-differential system of the order $n$, to one of the auxiliary system of the order $n+m$, where $m$ is a natural number,  which is “easily” for this analysis in a corresponding sense. Results for this auxiliary systems allow us to make conclusions for the given integro-differential system of the order $n$. We concentrate our attempts in the analysis of the distributed control in an integral form. An idea of reducing integro-differential systems to systems of ordinary differential equations is developed. We present results about the exponential stability of stationary points of integro-differential systems, using the method based on the presentation of solution with the help of Cauchy matrix. Various properties of integro-differential systems are studied by this way. Methods of general theory of functional differential equations developed by N.V.Azbelev and his followers are used. One of them is the Azbelev $W$-transform. We propose ideas allowing to achieve faster convergence to stationary point using a distributed control. We obtain estimates of solutions, using estimates of the Cauchy matrices.
\end{abstract}

\maketitle

\section{Introduction}
\label{sec:1}
Mathematical models in the form of systems of nonlinear ordinary differential equations, are used in many fields of science and technology to describe various phenomena. In medicine the purpose of mathematical modeling is the analysis and prediction of the development of diseases and their possible treatment. A comprehensive work on mathematical models in the field of immunology was summarized by Marchuk in his book \cite{12M1997}. The models constructed there reflect the most significant patterns of the immune system acting during these diseases. These model was studied in many works. Note, for example, the recent papers \cite{44M2004}, \cite{45MAG2015} and the bibliography therein. The adding control was proposed, for example, in \cite{9DBV2019}, \cite{16RC2014}, \cite{17S2017}, \cite{52BCDRV2019}, \cite{53DVPB2019}, \cite{54DVB2019},  \cite{55DVPS2017}, \cite{57DVP2019}, \cite{48B1986}. In the works \cite{7C2014},\cite{15RC2012} the basic mathematical model that takes into account the discrete control of the immune response is proposed. See also the recent papers \cite{9DBV2019}, \cite{52BCDRV2019}, \cite{54DVB2019}, where distributed control was considered. It can be noted that the use of information about behavior of a disease and the immune system for a long time (defined by distributed control, for example, in the form of an integral term) looks very natural in choosing strategy of a possible treatment. Optimal control in the basic model of the infection disease was considered in the work \cite{7C2014}, where the control function characterizing realization of an immunotherapy which includes in administration of immunoglobulin or donor antibodies is proposed. In the work  \cite{46AM1982}, the model of influence of an immunotherapy on dynamics of an immune response which represents generalization of basic model was considered. On the basis of the proposed model, the problem of determination of coefficients on the basis of laboratory dates was considered and a suitable management was proposed in  \cite{7C2014}, \cite{47B1979}. Such task was called control in uncertain conditions \cite{16RC2014}. A control algorithm in the uncertain conditions was proposed in the work (\cite{7C2014},  see pages 71-73).

In the recent papers \cite{9DBV2019}, \cite{52BCDRV2019}, \cite{54DVB2019} we present new approach for the study of the model of infection diseases. In this paper we summarize their results and formulate mathematical problems which look very natural from the medical point of view.

 Our contribution in the modeling is a distributed feedback control which is added to the equation describing the concentration of antibodies. This step transforms these systems to functional differential ones. As a result, we have to study the properties of solutions of these systems such as asymptotic behavior in the neighborhood of stationary points and stability of the stationary points. Importance of stationary points should be stressed. These points describe the conditions of the healthy body or the chronic disease. The aim of the treatment is to lead the process to one of the stationary points. Further we try to obtain estimates of solutions of linear and nonlinear systems of functional differential equations. One of the ways to these estimates is construction of the Cauchy matrix. First steps in this direction were proposed in the recent paper \cite{54DVB2019}.

\section{Description of model}
\label{sec:2}
In this paper we deal with the system of functional differential equations

$$
x^{\prime}(t)+(Ax)(t)=(\Phi Tx)(t),t\in [0,\infty),x=col\{x_1,…,x_n\} \eqno(2.1)
$$

where the operators $T$ and $A$ are linear continuous. $T,A:C_{[0,\infty)}^{n} \rightarrow{} L_{\infty[0,\infty)}^{n}$  $(C_{[0,\infty)}^{n}, L_{\infty[0,\infty)}^{n}$, are the spaces of continuous, and essentially bounded vector functions $x:[0,\infty) \rightarrow{} R^n$ respectively), $F:L_{\infty [0,\infty)}^{n} \rightarrow{} L_{\infty[0,\infty)}^n$ can be a linear or nonlinear bounded operator. 
We could analyze various boundary value problems for equation (2.1). One of them is the initial value problem. One of the main questions is the stability of this system \cite{4AS2002}. We consider the stationary points for corresponding operators in the spaces of continuous functions $C_{[0,\infty)}^n$ or essentially bounded functions $L_{\infty[0,\infty)}^n$.
We use our theoretical results in application to Marchuk’s model of infection diseases. This model reflects the most significant patterns of the immune system functioning during infectious diseases and focuses on the interactions between antigens and antibodies at different levels. We try to investigate the stability of stationary points of the immune system and its response to the treatment. We propose the control in the distributed form and obtain stabilization in the neighborhood of the stationary point in the model of infection diseases.
From the applications' point of view, the goal of the control in the system can be interpreted as a possibility to provide a corresponding immune response. It is noted in \cite{16RC2014} that the immune response mechanisms provides a key to understanding disease processes and methods of effective medical treatment \cite{12M1997}.
We try to combine our theoretical results with possible applications. Let us start with a description of one of these applications. Consider, for example, the Marchuk model of infection diseases:
$$
\left\{ 
\begin{array}{c}
\frac{dV}{dt}=\beta V\left( t\right) -\gamma F\left( t\right) V\left(
t\right) \\ 
\frac{dC}{dt}=\zeta \left( m\right) \alpha F\left(t\right)
 V\left(t\right) -\mu_{c}\left( C\left( t\right) -C^{\ast }\right) \\ 
\frac{dF}{dt}=\rho C\left( t\right)-\eta \gamma  F\left( t\right)  V\left(
t\right) -\mu _{f}F\left( t\right) \\ 
\frac{dm}{dt}=\sigma V\left( t\right) -\mu _{m}m\left( t\right)
\end{array} \right.  \eqno(2.2) 
$$
where
$V(t)$ – the antigen concentration rate,
$C(t)$ - the plasma cell concentration rate,
$F(t)$ - the antibody concentration rate,
$m(t)$ – the relative features of the body.
It is clear that system (2.2) can be presented in the form of general system (2.1).
Let us describe the coefficients: 
$\beta$ - coefficient describing the antigen activity,
$\gamma$ - the antigen neutralizing factor,
$\mu_f$  - coefficient inversely proportional to the decay time of the antibodies,
$\mu_m$  - coefficient inversely proportional to the organ recovery time,
$\mu_c$  -coefficient of reduction of plasma cells due to aging (inversely proportional to the lifetime),
$\sigma$ - constant related with a particular disease,
$\rho$ - rate of production of antibodies by one plasma cell.
Denote $C^*$  and $F^*$- the plasma rate concentration and antibody concentration of the healthy body, respectively.
It is assumed that during a certain period of time $\tau$, the plasma is restored as a result of the interaction between the antigen and the antibody cells. The product  $\zeta \left( m\right) \alpha F\left(t\right)
 V\left(t\right)$ includes the following coefficients:  $\alpha$ is the stimulation factor of the immune system.
The function
$$\zeta \left( m\right)={ 
\begin{array}{c}
1, \ \ 0 \leq m <m^* \\
\frac{1-m}{1-m^*}, \ \ m^*\leq m \leq 1
\end{array}  } ,
$$
is a continuous function, characterizing the health of the organ, which depends on the relative characteristics $m$ of the body, where $m^*$ is the maximum proportion of cells destroyed by antigens in the case that the normal functioning of the immune system is still possible.
This function is non-negative and does not increase. The function $m(t)$   can be described as  $1-\frac{1-M(t)}{1-M^* (t)}$, where $M(t)$ is the characteristic of a healthy organ (mass or area) and $M^*(t)$ is the corresponding characteristic of a healthy part of the affected organ.
Let us discuss now every equation in the model (2.2) in more detail form. The first equation
$\frac{dV}{dt}=\beta V\left( t\right) -\gamma F\left( t\right) V\left(
t\right)$  
presents the block of the virus dynamics. It describes the changes in the antigen concentration rate and includes the amount of the antigen in the blood. The antigen concentration decreases as a result of the interaction with the antibodies. The immune process characterizes the antibodies, whose concentration changes with time (destruction rate), is described by the equation:
$\frac{dF}{dt}=\rho C\left( t\right)-\eta \gamma  F\left( t\right)  V\left(
t\right) -\mu _{f}F\left( t\right)$.
The amount of the antibody cells also decreases as a result of the natural destruction. However, the plasma restores the antibodies and therefore the plasma state plays an important role in the immune process. Thus, the change in concentration rate of the plasma cell is included in several differential equations describing this system. Taking into account the healthy body level of plasma cells and their natural aging, the term $\mu_{c}(C(t)-C^* )$  is included in the second equation of system (2.2). The second and third equations present the humoral immune response dynamics.
Concerning the last equation of system (2.2):
$\frac{dm}{dt}=\sigma V\left( t\right) -\mu_{m}m\left( t\right)$.
The following can be noted 1) the value of $m$ increases with the antigen's concentration rate $V(t)$; 2) the maximum value of $m$ is unity, in the case of $100\%$ organ damage or  zero for a fully healthy organ. The coefficient $\mu_m$ describes the rate of generation of the target organ.
This model was considered in the recent work of Skvortsova \cite{17S2017}. Adding the control in the model introduced in Marchuk's book \cite{12M1997} is proposed, for example, in the works by Rusakov and Chirkov \cite{15RC2012},\cite{16RC2014}  where the importance of this development is explained.

\section{Stabilization through a support of the immune system}
\label{sec:3}
Our first goal is in stabilization of the process in the neighborhood of a suitable stationary solution. We make a corresponding linearization and then use the concepts of the stability theory proposed by N.V.Azbelev and his followers in the well-known books \cite{2ABBD2012},\cite{3AMR1991},\cite{4AS2002} for linear functional differential systems. The main idea is to choose “close” in a corresponding sense auxiliary linear system, to solve it and to construct its Cauchy matrix (see, for example \cite{52BCDRV2019}, \cite{54DVB2019}, 
\cite{56DVPS2016}). Then the scheme of the Azbelev $W$-transform is used. We propose new ideas in choosing “close” systems. For system of the order $n$, a corresponding  “close” system can be of the order $n+m$. Our main idea here is to reduce the analysis of a given system of the order $n$ to one of the auxiliary  system of the order $n+m$, which is “easily” in a corresponding sense. Results for the auxiliary systems allows us to make conclusions for the given system of the order $n$. We essentially concentrate our attempts in the analysis of the distributed control in an integral form. 
The integral terms reflect an orientation on average values in the construction of the control. Another reason of appearance of the integral terms is in the use of the "history of the process" to choose a strategy of a possible treatment. In our model, we demonstrate among other ideas that observation on the process of diseases can be very important in a treatment. It should be also noted that the proposed control can be realized practically. To sum up all these consequences, we can conclude that the control in the integral form is reasonable from the medical point of view.
Stability properties of integro-differential systems are studied.

Modifying model $(2.2)$, we propose the control in the following form
$$
u\left( t\right) =-b\int_{0}^{t}\left( F ( s\right) -F^{\ast} - \epsilon) e^{-k( t-s) } ds.  \eqno(3.1) 
$$

Adding this control in the third equation of $(2.2)$, we obtain the following system

$$
\begin{array}{c}
\frac{dV}{dt}=\beta V\left( t\right) -\gamma F\left( t\right) V\left(
t\right) \\ 
\frac{dC}{dt}=\zeta \left( m\right) \alpha F\left(t\right)
 V\left(t\right) -\mu_{c}\left( C\left( t\right) -C^{\ast }\right) \\ 
\frac{dF}{dt}=\rho C\left( t\right)-\eta \gamma  F\left( t\right)  V\left(
t\right) -\mu _{f}F\left( t\right)+u(t) \\ 
\frac{dm}{dt}=\sigma V\left( t\right) -\mu _{m}m\left( t\right)\\
\end{array} \eqno(3.2) 
$$
where $u(t)$ is defined by $(3.1)$.
Let $F^{*}$ be the value of the antibody concentration rate for a healthy body. While the case of $F^{*}>\frac{\beta}{\gamma}$ is considered by G.I. Marchuk in the book \cite{12M1997}. We try to consider the "bad" case where $F^{*}<\frac{\beta}{\gamma}$. It is clear that system (2.2) could not be stable in this case in the neighborhood of the stationary point $(0,C^{*},F^{*},0)$.
Consider the following system of five equations

$$
\begin{array}{c}
\frac{dV}{dt}=\beta V\left( t\right) -\gamma F\left( t\right) V\left(
t\right) \\ 
\frac{dC}{dt}=\zeta \left( m\right) \alpha F\left(t\right)
 V\left(t\right) -\mu_{c}\left( C\left( t\right) -C^{\ast }\right) \\ 
\frac{dF}{dt}=\rho C\left( t\right)-\eta \gamma  F\left( t\right)  V\left(
t\right) -\mu _{f}F\left( t\right)+u(t) \\ 
\frac{dm}{dt}=\sigma V\left( t\right) -\mu _{m}m\left( t\right)\\
\frac{du}{dt}=-b(F(t)-F^*-\varepsilon)-ku(t)
\end{array}. \eqno(3.3) 
$$

The following assertion allows us to reduce analysis of system $(3.2)$ to one of system $(3.3)$. 

\textbf{Lemma 3.1.}
\textit{The components of the solution-vector $y(t)=col( {v\left( t\right),s\left( t\right),f\left( t\right),m\left( t\right)})$ of system $(3.2)$ and four first components of the solution-vector $x(t)=col ({v\left( t\right),s\left( t\right),f\left( t\right),m\left( t\right),\tilde u\left( t\right)})$ of system $(3.3)$ satisfying the initial condition $ u\left( 0\right)=0$ coincide.}

\textbf{Theorem 3.1.}\textit{Let the inequality $\varepsilon \gamma > \beta - \gamma F^*$, $k>0$, $b>0$ be fulfilled, then the stationary solution $(0, C^*, F^*+\varepsilon,0,0)$ of system $(3.3)$ is exponentially stable}.\\ \\
To prove Theorem 3.1, we reduce the analysis of system (3.2) to one of system (3.3) by Lemma 3.1, linearize in the neighborhood of the stationary point


and then the negativity of roots to the characteristic polynomial of system (3.3) is demonstrated (see, for example, \cite{9DBV2019}).

Thus, we can stabilize the process at the point $(0, C^*, F^*+\varepsilon,0,0)$. It means that we have to support the immune system for a long time and to hold it on the level $F^*+\varepsilon$, where $\varepsilon > \frac{\beta - \gamma F^{*}}{\gamma}$.

\section{Distributed Control and the Lyapunov Characteristic Exponents}
\label{sec:4}
To use the control in order to make convergence to set stationary state faster is the second goal. Note that the stationary points present the condition of the healthy body or at least chronical process of disease which we try to reach. This problem is directly related to the duration of a possible treatment. In many cases, this may have an important influence on the choice of treatment method and on the decision on the acceptability of such treatment in principle.

The goal of this part to obtain faster tending to set stationary state.

Consider the system
$$
\begin{array}{c}
\frac{dv}{dt}=\beta v\left( t\right) -\gamma F^{\ast} f\left(
t\right)v\left(t\right) \\ 
\frac{ds}{dt}=\alpha V_{m} \frac{F^{\ast}}{C^{\ast }}\zeta \left(m\right) f\left( t \right) v\left(t\right)-\mu _{c}\left(s\left( t\right) -1\right) \\ 
\frac{df}{dt}=\frac{\rho C^{\ast}}{F^{\ast}}s\left( t\right)-\eta \gamma V_{m} f\left( t\right) v\left( t\right) -\mu_{f} f\left( t\right)-b\tilde {u}\left( t\right) \\ 
\frac{dm}{dt}=\sigma V_{m} v\left( t\right) -\mu _{m}m\left( t\right)\\
\frac{d\tilde{u}}{dt}=f\left(t\right)-1 -k\tilde {u} \left(t\right)%
\end{array} \eqno(4.1)
$$

where 

$$
\tilde{u}=\int_{0}^{t}(f(s)-1)e^{-k(t-s)}ds
$$.

Denoting in $(4.1)$  
$$\alpha_{1}=\beta, \alpha_{2}= \gamma F^{\ast} , \alpha_{3} =\alpha V_{m} \frac{F^{\ast}}{C^{\ast }}, \alpha_{4}=\mu_{f}=\frac{\rho C^{\ast}}{F^{\ast}},\alpha_{5}=\mu
_{c} ,\alpha_{6}=\sigma V_{m},\alpha_{7}=\mu
_{m}, \alpha_{8}=\eta \gamma V_{m}, \eqno(4.2)
$$
and linearizing system $(4.1)$ in the neighborhood of stationary point $v=m=\tilde{u}=0$, $s=f=1$, we can write system $(4.1)$ in the form

$$
\begin{array}{c}
\frac{dx_{1}}{dt}=(\alpha_{1} -\alpha_{2})x_{1}\\ 
\frac{dx_{2}}{dt}=\alpha_{3}   x_{1} -\alpha_{5}x_{2} \\ 
\frac{dx_{3}}{dt}=-\alpha_{8}x_{1}+ \alpha_{4}x_{2}  -\alpha_{4}x_{3}-bx_{5} \\ 
\frac{dx_{4}}{dt}=\alpha_{6} x_{1} -\alpha_{7}x_{4}\\
\frac{dx_{5}}{dt}=x_{3} -k x_{5}%
\end{array}, \eqno(4.3)
$$

and to linearize system $(2.2)$ and write it in the form 
$$
\begin{array}{c}
\frac{dx_{1}}{dt}=(\alpha_{1} -\alpha_{2})x_{1}\\ 
\frac{dx_{2}}{dt}=\alpha_{3}   x_{1} -\alpha_{5}x_{2} \\ 
\frac{dx_{3}}{dt}=-\alpha_{8}x_{1}+ \alpha_{4}x_{2}  -\alpha_{4}x_{3} \\ 
\frac{dx_{4}}{dt}=\alpha_{6} x_{1} -\alpha_{7}x_{4}\\
\end{array} \eqno(4.4)
$$
Denote $\lambda_{i},i=\{1,4\}$ the roots of the characteristic polynomial of systems $(4.4)$, and $\tilde{\lambda}=\max_{1 \leq i \leq 4} \lambda_{i}, \tilde{\lambda^*}=\max_{1 \leq j \leq 5} \mathbf{Re}(\lambda_{j}^*$) of $(4.3)$.\\

\textbf{Theorem 4.1.} 
\textit{If $\beta<\gamma F^*,$ $b>0$ and $k>0$, then integro-differential system $(4.3)$ is exponentially stable and if in addition the inequality $k> \alpha_4$ is fulfilled then  $\tilde{\lambda} \geq \tilde{\lambda^*}.  $}\\ \\
To prove Theorem 4.1, after reducing analysis of system (3.2) to one of system (3.3) by Lemma 3.1 and linearizing in the neighborhoods of the stationary points  of system (2.2) and (3.3) respectively, we compare the roots of characteristic polynomials of system (4.3) and (4.4) (see, for example \cite{52BCDRV2019}).\\

On Figures 1-4 the solution of model of the pneumonia with the natural flow of data without the control of disease are presented by curves of red color, disease in the case of considered distributed control-by curves of green color. 

Figure 1 demonstrates the dynamics  in antigen concentration during the course of the disease.
The insert detailing  the process in the first two days was performed on a different scale and demonstrates  the fact that the management transfers the disease from the “acute” form to the “subclinical” one (the antigen concentration only decreases after injetion).
Figure 2 demonstrates the dynamics in plasma cell concentration during the disease process.
It can be seen from the figure that control leads to a faster increase in the concentration of plasma cells, which in this case ensures a transition to the “subclinical” form of the disease. In addition, it is necessary to note a fourfold increase in the maximum concentration of plasma cells in the case of control, compared with the option without control.
Figure 3 demonstrates the dynamics in antibody concentration during the disease process.
The graph shows that the concentration of antibodies in the solution with control practically does not change, because in this case they are replaced by donor antibodies, which is what the control actually consists of.
The dynamics in the proportion of target organ cells destroyed by antigen during the disease process is presented on Figure 4. The values for the variant with control are given with an increase of $10^4$ times.
Thus, control allows to reduce the maximum proportion of affected cells of the target organ by more than $2.5 \times 10^4$ times.

\begin{figure}
   \centering
       \includegraphics[width=\textwidth]{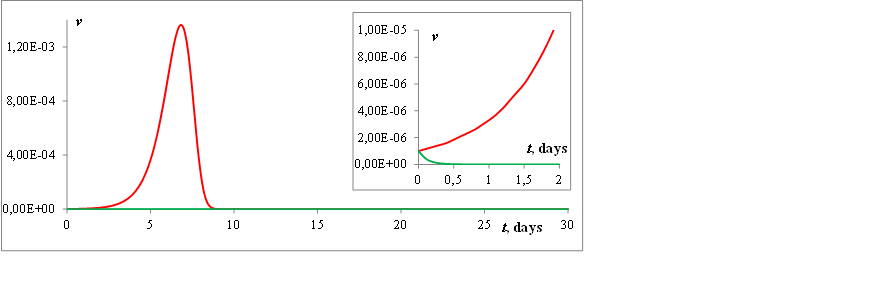}
   \caption{Dynamics of the immune response: antigen}
\end{figure}

\begin{figure}
    \centering
        \includegraphics[width=\textwidth]{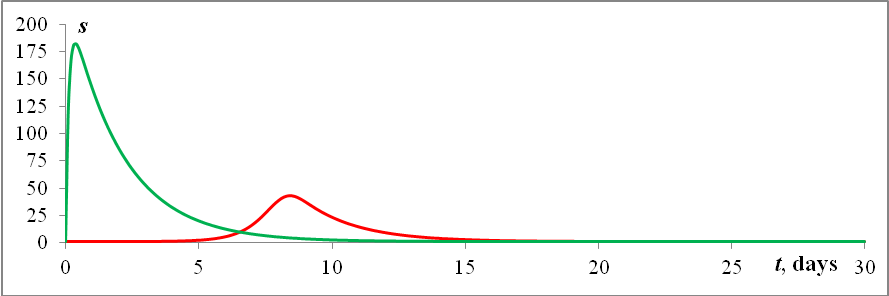}
    \caption{Dynamics of the immune response: plasma}
\end{figure}

\begin{figure}
    \centering
       \includegraphics[width=\textwidth]{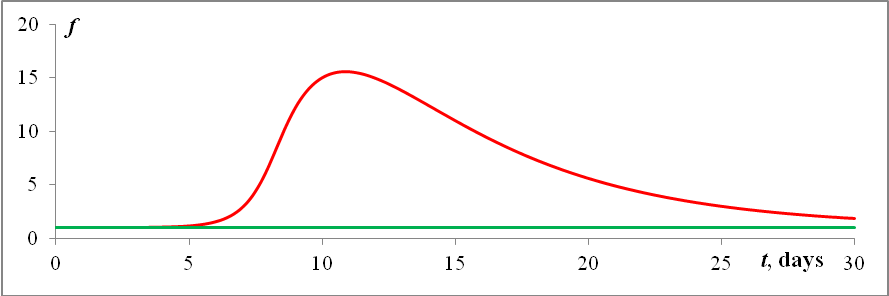}
    \caption{Dynamics of the immune response: antibodies}
\end{figure}

\begin{figure}
    \centering
        \includegraphics[width=\textwidth]{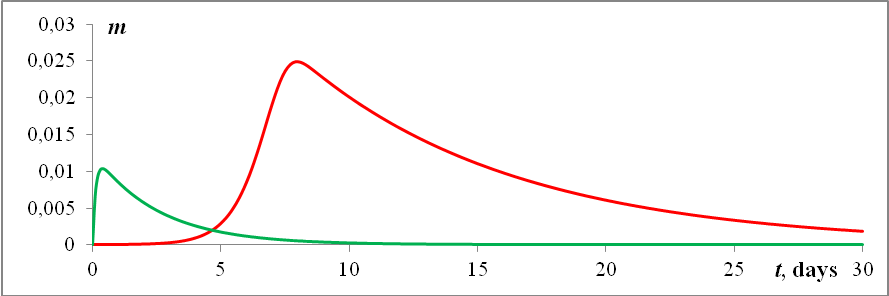}
   \caption{Dynamics of the immune response: rate of the destroyed cells}
\end{figure}

\section{Cauchy Matrix}
\label{sec:5}

To estimate the size of the neighborhood of the stationary solutions which usually describe the states of the healthy body is the third goal of our research. In practical problems it is necessary since we have to hold process in a corresponding zone. Process going beyond a certain admissible neighborhood of a stationary solution may be dangerous for patients. 

In constructing every model, the influences of various additional factors
that have seemed to be nonessential were neglected. The influence effect of
choosing nonlinear terms by their linearization in neighborhood of
stationary solution is also neglected. Even in the frame of linearized
model, only approximate values of coefficients instead of exact ones\ are
used. Changes of these coefficients with respect to time are not usually taken into
account. It looks important to estimate an influence of all these factors.

In order to make this we have to obtain estimates of the elements of the Cauchy matrix of corresponding linearized (in a neighborhood of a stationary point) system.
Consider the system 
$$
x'(t)=P(t)x(t)+G(t),
$$
where $P(t)$ is a $(n \times n)$-matrix, $G(t)$ is $n$-vector. Its general solution $x(t)=col\{x_1(t),...x_n(t)\}$ can be represented in the form (see, for example, \cite{2ABBD2012})
$$
x(t)=\int_0^t C(t,s) G(s) ds + C(t,0)x(0),
$$
where $n \times n$-matrix $C(t,s)$ is called the Cauchy matrix. Its $j$-th column ($j=1,...,n$) for every fixed $s$ as a function of $t$, is a solution of the corresponding homogeneous system
$$
x'(t)=P(t)x(t),
$$
satisfying the initial conditions $x_i(s)=\delta_{ij},$ where

\[
\delta _{ij}=\left\{ 
\begin{array}{c}
1,\text{ \ \ }i=j, \\ 
0,\text{ \ \ }i\neq j,%
\end{array}%
\right. \text{ \ }i=1,...,n,
\]
This Cauchy matrix $C(t,s)$ satisfies the following symmetric properties $C(t,s)=X(s)X^{-1}(s),$ where $X(t)$  is a fundamental matrix, $C(t,0)=C(t,s)C(s,0)$, and in the case of constant matrix $P(t)=P,$ we have $X(t-s)=C(t,s)$ is a fundamental matrix for every $s \geq 0.$
These definition and properties allow us to construct and estimate $C(t,s)$.
The construction of the Cauchy matrix of system $(4.3)$ can be found, for example, in \cite{54DVB2019}.

\section{Stabilization with the use of uncertain coefficient in the control}
\label{sec:6}

Consider the following system of equations with uncertain coefficient in the control

$$
\begin{array}{c}
\frac{dV}{dt}=\beta V\left( t\right) -\gamma F\left( t\right) V\left(
t\right) \\ 
\frac{dC}{dt}=\zeta \left( m(t)\right) \alpha F\left( t\right) V\left( t\right)
-\mu _{c}\left( C\left( t\right) -C^{\ast }\right) \\ 
\frac{dF}{dt}=\rho C-\eta \gamma F\left( t\right) V\left( t\right) -\mu
_{f}F\left( t\right) -\left( b+\bigtriangleup b(t)\right) u\left( t\right)
\\ 
\frac{dm}{dt}=\sigma V\left( t\right) -\mu _{m}m\left( t\right) \\ 
\frac{du}{dt}=F\left( t\right) -F^{\ast }-ku\left( t\right)%
\end{array} \eqno(6.1) 
$$

where
$u(t) = \int\limits_{0}^{t}\left( F\left( s\right) -F^{\ast
}\right) e^{-k\left( t-s\right) }ds $

This system can be rewritten in the form

$$
\left\{ 
\begin{array}{c}
x_{1}^{\prime }=\left( a_{1}-a_{2}\right) x_{1}+g_{1}\left( x_1(t), x_3(t)\right) \\ 
x_{2}^{\prime }=a_{3} x_{1}-a_{5}x_{2}+g_{2}\left(  x_1(t), x_3(t)\right) \\ 
x_{3}^{\prime }=-a_{8}\ x_{1}+a_{4}x_{2}-a_{4}x_{3}-\left( b+\bigtriangleup
b(t)\right) x_{5}+g_{3}\left( x_1(t), x_3(t)\right) \\ 
x_{4}^{\prime }=a_{6}x_{1}-a_{7}x_{4} \\ 
x_{5}^{\prime }=x_{3}-kx_{5}%
\end{array}%
\right. ,\eqno(6.2) 
$$
where $g_{i}(x_1(t), x_3(t))\left( t\right) ,$ \ $1\leq i\leq 3$ \ results of "mistakes" we
made in the process of the linearization.\\
Consider the system
$$
X^{\prime }=AX+\Delta B\left( t\right) X+F\left( t\right) ,\eqno(6.3) 
$$

where

\begin{eqnarray*}
X(t) &=&\left( 
\begin{array}{c}
x_{1}(t) \\ 
x_{2}(t) \\ 
x_{3}(t) \\ 
x_{4}(t) \\ 
x_{5}(t)%
\end{array}%
\right) ,\text{ \ \ \ }\Delta B\left( t\right) =\left( 
\begin{array}{ccccc}
0 & 0 & 0 & 0 & 0 \\ 
0 & 0 & 0 & 0 & 0 \\ 
0 & 0 & 0 & 0 & -\bigtriangleup b(t) \\ 
0 & 0 & 0 & 0 & 0 \\ 
0 & 0 & 0 & 0 & 0%
\end{array}%
\right). \\
\end{eqnarray*}

On the basis of the estimates of the elements of the Cauchy matrix we obtain the following assertions on the stability of system (6.2).
Denoting $Q_{j}=ess \sup _{t \geq 0} \int\limits_{0}^{t}\sum\limits_{i=1}^{5}\left\vert
\left( \Delta B\left( t\right) C(t,s)\right) _{ij}\right\vert ds$ and $\bigtriangleup b^{\ast }=ess\sup_{t\geq 0}\left\vert
\bigtriangleup b(t)\right\vert $, we obtain the estimates:

$$
\begin{array}{c}
Q_{1}\leq \bigtriangleup b^{\ast }\left[ 
\begin{array}{c}
\left\vert \frac{\alpha _{24}\left( \alpha _{32}-\alpha _{35}\right) -\alpha
_{25}\left( \alpha _{32}-\alpha _{34}\right) }{\alpha _{15}\alpha
_{24}\left( \alpha _{31}-\alpha _{32}\right) }\right\vert \frac{1}{%
\left\vert \lambda _{1}\right\vert }+ \\ 
\left\vert \frac{\alpha _{24}\left( \alpha _{31}-\alpha _{35}\right) -\alpha
_{25}\left( \alpha _{31}-\alpha _{34}\right) }{\alpha _{15}\alpha
_{24}\left( \alpha _{31}-\alpha _{32}\right) }\right\vert \frac{1}{%
\left\vert \lambda _{2}\right\vert }+\left\vert \frac{\alpha _{25}}{%
a_{5}\alpha _{15}\alpha _{24}}\right\vert%
\end{array}%
\right] , \\ 
\\ 
Q_{2}\leq \bigtriangleup b^{\ast }\left[ \left\vert \frac{\alpha
_{32}-\alpha _{34}}{\alpha _{24}\left( \alpha _{31}-\alpha _{32}\right) }%
\right\vert \frac{1}{\left\vert \lambda _{1}\right\vert }+\left\vert \frac{%
\alpha _{31}-\alpha _{34}}{\alpha _{24}\left( \alpha _{31}-\alpha
_{32}\right) }\right\vert \frac{1}{\left\vert \lambda _{2}\right\vert }+%
\frac{1}{\left\vert a_{5}\alpha _{24}\right\vert }\right] , \\ 
\\ 
Q_{3}\leq \bigtriangleup b^{\ast }\left[ \frac{1}{\left\vert \alpha
_{31}-\alpha _{32}\right\vert }\frac{1}{\left\vert \lambda _{1}\right\vert }+%
\frac{1}{\left\vert \alpha _{31}-\alpha _{32}\right\vert }\frac{1}{%
\left\vert \lambda _{2}\right\vert }\right] , \\ 
\\ 
Q_{4}=0, \\ 
\\ 
Q_{5}\leq \bigtriangleup b^{\ast }\left[ \left\vert \frac{\alpha _{32}}{%
\alpha _{31}-\alpha _{32}}\right\vert \frac{1}{\left\vert \lambda
_{1}\right\vert }+\left\vert \frac{\alpha _{31}}{\alpha _{31}-\alpha _{32}}%
\right\vert \frac{1}{\left\vert \lambda _{2}\right\vert }\right] .%
\end{array}%
\eqno(6.4) 
$$

\textbf{Theorem 6.1. } \cite{54DVB2019} \textit{Let $k>0$, $b>0$ and $a_{i},$ $1\leq i\leq 8$, are
real positive and different, $a_1 < a_2$,  }$\left( a_{4}-k\right) ^{2}>4b$ \textit{and the
inequality }$\max_{1\leq j\leq 5}\left\{ \left\vert Q_{j}\right\vert
\right\} <1$ \textit{be true. Then system }$(6.2)$ \textit{is exponential
stable.}\\
Denoting $P_{j}=ess \sup _{t \geq 0} \int\limits_{0}^{t}\sum\limits_{i=1}^{5}%
\left\vert \left( \Delta B\left( t\right) C(t,s)\right) _{ij}\right\vert ds$,
we obtain the estimates

$$
\begin{array}{c}
P_{1}\leq \bigtriangleup b^{\ast }\left[ 
\begin{array}{c}
\left\vert \frac{\beta _{24}\beta _{35}-\beta _{25}\beta _{34}}{\beta
_{31}\beta _{15}\beta _{24}}\right\vert \frac{4}{\left(
a_{4}+k\right)^2 }+\left\vert \frac{\beta _{24}\left( \beta _{31}-\beta
_{35}\right) -\beta _{25}\left( \beta _{31}-\beta _{34}\right) }{\beta
_{31}\beta _{52}\beta _{24}\beta _{15}}\right\vert \frac{4}{\left\vert
a_{4}^{2}-k^{2}\right\vert } \\ 
+\left\vert \frac{\beta _{25}}{\beta _{15}\beta _{24}}\right\vert \frac{1}{%
\left\vert a_{5}\right\vert }+\frac{1}{\left\vert \beta _{15}\right\vert }%
\frac{1}{\left\vert a_{1}-a_{2}\right\vert }%
\end{array}%
\right] , \\ 
\\ 
P_{2}\leq \bigtriangleup b^{\ast }\left[ \left\vert \frac{\beta _{34}}{\beta
_{24}\beta _{31}}\right\vert \frac{4}{\left( a_{4}+k\right)^2 }%
+\left\vert \frac{\beta _{31}-\beta _{34}}{\beta _{31}\beta _{24}\beta _{52}}%
\right\vert \frac{4}{\left\vert a_{4}^{2}-k^{2}\right\vert }+\frac{1}{%
\left\vert \beta _{24}\right\vert }\frac{1}{\left\vert a_{5}\right\vert }%
\right] , \\ 
\\ 
P_{3}\leq \bigtriangleup b^{\ast }\left[ \frac{1}{\left\vert \beta
_{31}\right\vert }\frac{4}{\left( a_{4}+k\right)^2 }+\frac{1}{%
\left\vert \beta _{31}\beta _{52}\right\vert }\frac{4}{\left\vert
a_{4}^{2}-k^{2}\right\vert }\right] , \\ 
\\ 
P_{4}=0, \\ 
\\ 
P_{5}\leq \bigtriangleup b^{\ast }\frac{1}{\left\vert \beta _{52}\right\vert 
}\frac{4}{\left\vert a_{4}^{2}-k^{2}\right\vert }.%
\end{array}%
\eqno(6.5) 
$$

\textbf{Theorem 6.2. } \cite{54DVB2019} \textit{Let $k>0$, $b>0$ and $a_{i},$ $1\leq i\leq 8$, are
real positive and different, $a_1 < a_2$,  }$\left( a_{4}-k\right) ^{2}=4b$ \textit{and the
inequality }$\max_{1\leq j\leq 5}\left\{ \left\vert P_{j}\right\vert
\right\} <1$ \textit{be true. Then system }$(6.2)$ \textit{is exponential
stable.}

Denoting $R_{j}=ess \sup _{t \geq 0} \int\limits_{0}^{t}\sum\limits_{i=1}^{5}%
\left\vert \left( \Delta B\left( t\right) C(t,s)\right) _{ij}\right\vert ds$
we obtain estimates

$$
\begin{array}{c}
R_{1}\leq \bigtriangleup b^{\ast }\left[ 
\begin{array}{c}
\left\vert \frac{\gamma _{24}-\gamma _{25}}{\gamma _{15}\gamma _{24}}%
\right\vert \frac{2}{\left\vert a_{4}+k\right\vert }+\left\vert \frac{\gamma
_{24}\left( 2\gamma _{35}-a_{4}+k\right) +\gamma _{25}\left(
a_{4}-2a_{5}+3k\right) }{\gamma _{32}\gamma _{15}\gamma _{24}}\right\vert 
\frac{1}{\left\vert a_{4}+k\right\vert } \\ 
+\left\vert \frac{\gamma _{25}}{\gamma _{15}\gamma _{24}}\right\vert \frac{1%
}{\left\vert a_{5}\right\vert }+\left\vert \frac{1}{\gamma _{15}}\right\vert 
\frac{1}{\left\vert a_{1}-a_{2}\right\vert }%
\end{array}%
\right] , \\ 
\\ 
R_{2}\leq \bigtriangleup b^{\ast }\left[ \frac{1}{\left\vert \gamma
_{24}\right\vert }\frac{2}{\left\vert a_{4}+k\right\vert }+\left\vert \frac{%
a_{4}-3k+2a_{5}}{\gamma _{24}\gamma _{32}}\right\vert \frac{1}{\left\vert
a_{4}+k\right\vert }+\frac{1}{\left\vert \gamma _{24}\right\vert }\frac{1}{%
\left\vert a_{5}\right\vert }\right] , \\ 
\\ 
R_{3}\leq \bigtriangleup b^{\ast }\frac{1}{\left\vert \gamma
_{32}\right\vert }\frac{2}{\left\vert a_{4}+k\right\vert }, \\ 
\\ 
R_{4}=0, \\ 
\\ 
R_{5}\leq \bigtriangleup b^{\ast } \left[ \frac{2}{\left\vert a_4 + k\right\vert} +  \left\vert \frac{a_{4}-k}{\gamma _{32}}%
\right\vert \frac{1}{\left\vert a_{4}+k\right\vert }\right].%
\end{array}%
\eqno(6.6) 
$$

\textbf{Theorem 6.3. } \cite{54DVB2019} \textit{Let $k>0$, $b>0$ and $a_{i},$ $1\leq i\leq 8$, are
real positive and different, $a_1 < a_2$,  }$\left( a_{4}-k\right) ^{2}<4b$ \textit{and the
inequality }$\max_{1\leq j\leq 5}\left\{ \left\vert R_{j}\right\vert
\right\} <1$ \textit{be true. Then system }$(6.2)$ \textit{is exponential
stable.}\\\\

\textbf{Remark 6.1.} \textit{Note that the approach presented here can be used in the model of testosterone regulation (see, for example, \cite{53DVPB2019}, \cite{AIP}, \cite{DVS}, \cite{DV}).}\\\\

\textbf{Acknowledgements}
This paper is part of the third and fourth author’s Ph.D. thesis which is being carried out in the Department of Mathematics at Ariel University.


\begin{thebibliography}{99}

\bibitem{1ABDG2005}  R. Agarwal, M. Bohner, A. Domoshnitsky, Ya. Goltser, Floquet theory and stability of nonlinear integro-differential equations, Acta Mathematica Hungarica 109 (4), 305-330, 2005.

\bibitem{2ABBD2012} R. Agarwal, L. Berezansky, E. Braverman and A. Domoshnitsky, Nonoscillation Theory of Functional Differential Equations with Applications, Springer, 2012.

\bibitem{3AMR1991}  N.V.Azbelev, V.P.Maksimov and L.F.Rakhmatullina, Introduction to theory of functional-differential equations, Nauka, Moscow, 1991.

\bibitem{4AS2002} N.V.Azbelev, P.M.Simonov, Stability of differential equations with aftereffect, Reference - 240 Pages, ISBN 9780415269575 - CAT TF1327, 2002.

\bibitem{7C2014}  M.V. Chirkov, Parameter Identification and Control in Mathematical Models of the Immune Response. Thesis. Perm State University. Perm. Russia,2014.

\bibitem{9DBV2019}  A. Domoshnitsky, M. Bershadsky, I.Volinsky, Distributed control in stabilization of model of infection diseases, Russian Journal of Biomechanics. 2019. Vol. 23, No. 4: 494-499.
DOI:10.15593/RJBiomech/2019.4.08

\bibitem{12M1997}  G.I. Marchuk, Mathematical Modelling of Immune Response in Infection Diseases, Mathematics and Its Applications, Springer, 1997.

\bibitem{15RC2012}   S.V.Rusakov, M.V.Chirkov, Mathematical model of influence of immunotherapy on dynamics of immune response, Problems of Control, No.6, pp. 45-50, 2012.

\bibitem{16RC2014}  S.V.Rusakov and M.V.Chirkov, Identification of parameters and control in mathematical models of immune response, Russian Journal of Biomechanics, vol.18, No.2, 259-269, 2014

\bibitem{17S2017} M.Skvortsova, Asymptotic Properties of Solutions in Marchuk's Basic Model of Disease, Functional Differential Equations, 2017, vol. 24, no 3-4, pp.127-135.

\bibitem{52BCDRV2019} M. Bershadsky, M. Chirkov, A. Domoshnitsky, S. Rusakov, and I. Volinsky, "Distributed Control and the Lyapunov Characteristic Exponents in the Model of Infectious Diseases", Complexity, Volume 2019, Article ID 5234854, Published 13 November 2019
https://doi.org/10.1155/2019/5234854

\bibitem{53DVPB2019} A. Domoshnitsky, I.Volinsky, O.Pinhasov, M.Bershadsky, Questions of Stability of Functional Differential Systems around the Model of Testosterone Regulation, Boundary Value Problems 2019 (1), 1-13; https://doi.org/10.1186/s13661-019-01295-2
\bibitem{54DVB2019} A. Domoshnitsky, I.Volinsky, M.Bershadsky,  Around the Model of Infection Disease: The Cauchy Matrix and Its Properties, Symmetry 2019, 11(8),1016; https://doi.org/10.3390/sym11081016 (Published: August 6,  2019)

\bibitem{55DVPS2017} Domoshnitsky A., Volinsky I., Polonsky A., Sitkin A.,
Stabilization by delay distributed feedback control, {\em Mathematical Modeling of Natural Phenomena} 12 (6), 91-105,  2017. 

\bibitem{56DVPS2016} A Domoshnitsky, I Volinsky, A Polonsky, A Sitkin, Practical constructing the Cauchy function of integro-differential equation, {\em FUNCTIONAL DIFFERENTIAL EQUATIONS 23} (3-4), 109–-118, 2016.

\bibitem{57DVP2019} Domoshnitsky A., Volinsky I., Polonsky A., Stabilization of third order differential equation by delay distributed feedback control,
{\em Math. Slovaca:  No. 5, Volume 69}, 1165-1175, 2019, DOI: https://doi.org/10.1515/ms-2017-0298.

\bibitem{AIP}Alexander Domoshnitsky, Irina Volinsky and Olga Pinhasov, Some developments in the model of testosterone regulation, AIP Conference Proceedings 2159, 030010 (2019); https://doi.org/10.1063/1.5127475

\bibitem{DVS} Alexander Domoshnitsky, Irina Volinsky, Roman Shklyar, About Green’s functions for impulsive differential equations {\em FUNCTIONAL DIFFERENTIAL EQUATIONS 20} (1-2), 55–-81, 2013.

\bibitem{DV} Alexander Domoshnitsky, Irina Volinsky, About differential inequalities for nonlocal boundary value problems with impulsive delay equations, {\em Mathematica Bohemica 140} (2), 121-128, 2015.

\bibitem{44M2004}  V. P. Martsenyuk, Construction and study of stability of an antitumor immunity model, Cybernetics and Systems Analysis, Vol. 40, No. 5, 2004

\bibitem{45MAG2015} V.P.Martsenyuk, I.Ye.Andrushchak and I.S.Gvozdetska, Qualitiative analysis of the antineoplastic immunitty system on the basis of a decision tree, Cybernetics and systems analysis, Vol. 51, No. 3, May, 2015 

\bibitem{46AM1982} Asachenkov A.L., Marchuk G.I. The specified mathematical model of an infectious disease, Mathematical modeling in immunology and medicine,under the editorship of G.I.Marchuk. Novosibirsk: Science, 1982, Page 44-59 (in Russian).

\bibitem{47B1979} Bard Y. Nonlinear estimation of parameters. Statistics, 1979 (in Russian). 

\bibitem{48B1986} Belih L.N. O the numerical solution of models of diseases//Mathematical models in immunology and medicine / under the editorship of G.I. Marchuk, L.N.Belih. - M.: World, 1986. - Page 291-297 (in Russian).



\end{thebibliography}
\end{document}